\documentstyle[aps,prb,floats,preprint,eqsecnum,tighten,epsfig]{revtex}
\begin{document}
\widetext
\thispagestyle{empty}

\title{Emergence of macroscopic temperatures in systems that 
  are not  thermodynamical microscopically: towards a thermodynamical description
of slow granular rheology.}

\author{Jorge Kurchan}
\address{ 
\it P.M.M.H. Ecole Sup\'erieure de Physique et Chimie Industrielles,
\\
10, rue Vauquelin, 75231 Paris CEDEX 05,  France}

\date\today
\maketitle
\widetext

\begin{abstract}

A  scenario for systems with slow dynamics is characterised by
 stating that there are  several temperatures coexisting in the
 sample, with a  single  temperature shared by  all observables
 at each (widely separate) time-scale.

 In preparation for the study of granular rheology, 
we show within this framework
 that glassy systems  with driving  and  friction that are generic and do not
correspond to a thermal bath --- and whose  microscopic `fast' motion is
hence not thermal ---
have   a  well-defined macroscopic 
temperature associated to the  slow degrees of freedom. 

This temperature is what a  thermometer coupled to the system will
measure if tuned to respond to low frequencies, and since it can be
related
to the number of stationary configurations, it is the formalisation
of Edwards' `compactivity' ideas.

\end{abstract}
\vspace{.5cm}
 
\section{}

 Granular matter set into motion by shearing, shaking or tapping is one of the 
most interesting cases of macroscopic out of equilibrium systems.
 Given a granular system subjected to some form of power input that makes it perform
stationary flow on average, a very natural question that arises is to what extent
it resembles a thermodynamic  system of interacting particles such as, 
for example, a liquid.

More specifically, many attempts have been made to define a 
`granular temperature' (see
e.g. \cite{Savage,Sam}).
In order to deserve its name, a temperature has to play the role of
 deciding the direction
of heat flow: it must be connected to a form of the {\em zero-th} law.
In order to pursue this line, however,  one has to somehow take care of
the characteristics of granular flow that distinguish it from usual kinetic theory:

1) Energy is not conserved, and, more generally, the motion 
does not have the very strong  phase-space volume conservation
properties typical of Hamilton's equations. 
The dissipation is due to  friction which is in general not linear in the velocity,
and dependent upon the relative positions of the particles.

2) Power  is supplied by tapping (which may be periodic in time) or by shearing,
a manner  very different from that  of the `collisions' of a thermal bath.

Under these circumstances, there is no reason why the observables should be related to
a Gibbs (or equivalent) ensemble, and the possibility of having
thermodynamic 
concepts seems lost.
 
In this paper we
  shall consider situations with shear and friction,
 in  the limit of weak shear. The effect of a coherent `tapping' will
be discussed in further work \cite{Berthier}.
In this limit  of `slow rheology',  it will turn out that even though
the rapid motion cannot be associated with thermal motion, there appears 
 for the slow flow a natural temperature playing the usual role in thermometry
 and thermalisation \cite{leticia}.

The computation presented here can be done in a wide range of
 approximation schemes consisting
in resumming  a perturbative expression for the dynamics in several 
forms --- and to higher and
higher levels of approximation --- in particular
the so-called mode-coupling approximation.  Here, for concreteness,
 I will carry it through for 
a simple model for which the (single mode) mode-coupling approximation is exact.

\vspace{.5cm}

{\bf Multiple Thermalisation in Aging and Rheology}

\vspace{.5cm}

Granular systems have been recognised as being closely related  to glassy 
systems \cite{Struick}. Accordingly, several recent developments and models have been 
borrowed from the field of glasses to understand their properties\cite{Sollich}. 

A picture has arisen in the last few years 
for aging or {\em gently} driven glasses involving multiple thermalisations at 
widely separated timescales (see \cite{review} for a review). 
In the simplest scheme, the situation  
is as follows: Given any two observables 
$A$ and $B$ belonging to the system, define the correlation
function as:
\begin{equation}
\langle A(t) B(t') \rangle = C_{AB}(t,t')
\end{equation}
and the response of $A$ to a field conjugate to $B$:
\begin{equation}
\frac{\delta}{\delta h_B(t')} \langle A(t)  \rangle = R_{AB}(t,t')
\end{equation}
For a pure relaxational (undriven) glass, the correlation breaks up into two parts:
\begin{equation}
C_{AB}(t,t') = C^F_{AB}(t-t') + {\tilde{C}}_{AB} \left( \frac{h(t')}{h(t)} \right)
\end{equation}
with $h$ the {\em same} growing function
 for all observables $A$, $B$. The fact that $C_{AB}(t,t')$ never
becomes a function of the time-differences 
means that the system is forever out of equilibrium,
it {\em ages}.
If instead the glass is  gently driven 
(with driving forces  proportional to, say, $\epsilon$),
 aging may stop, and we have:
\begin{equation}
C_{AB}(t,t') = C^F_{AB}(t-t') + {\tilde{C}}_{AB} \left( \frac{t-t'}{\tau_o} \right)
\end{equation}
 where $\tau_o$ is a time scale that diverges as $\epsilon$ goes to zero.

In the long time limit and in the small drive limit,
 the time scales become very separate.
When this happens, it turns out that the responses behave as:
\begin{equation}
 R_{AB}(t,t') = \beta \frac{\partial}{\partial t'}
 C^F_{AB}+ \beta^* \frac{\partial}{\partial t'}  {\tilde{C}}_{AB} 
\end{equation}
in the aging and the driven case.

The fast degrees of freedom behave as if thermalised at the bath temperature $\beta$. 
On the other hand, the effective, system-dependent  temperature $T^* = 1/ \beta^*$ 
indeed deserves its name: it can be shown \cite{Cukupe,statphys,jamming} that it is what a `slow'
 thermometer measures,
and it controls the heat flow and the thermalisation of the slow degrees of freedom.
It is {\em the same} for any two observables at a given timescale, whether the system is aging
or gently driven.
Furthermore, it is {\em macroscopic}: it remains non-zero in the limit in which the bath temperature is zero.

If the system is not coupled to a true thermal bath, but energy is supplied by shaking and shearing, while it
is dissipated by a nonlinear complicated friction, there is no bath temperature $\beta$.
What will be argued  in what follows is that even so, the `slow' temperature $\beta^*$ survives despite
the fact that the fast motion is not thermal in that case. Indeed, if we have  correlation having
fast and slow components:
 \begin{equation}
C_{AB}(t,t') = C^F_{AB}(t,t') + {\hat{C}}_{AB}^S (t,t')
\end{equation}
the response is of the form:
 \begin{equation}
R_{AB}(t,t') = R^F_{AB}(t,t') + \beta^* \frac{\partial}{\partial t'} {\hat{C}}_{AB}^S (t,t')
\end{equation}
with the fast response  $R^F_{AB}(t,t')$ bearing no general relation
with the fast correlation $C^F_{AB}(t,t') $. 

The effective `slow' temperature so defined   is then found to be
 directly related to Edwards'
compactivity \cite{Sam}, but, in the spirit of Ref. \cite{anita},
in the context of  slowly moving rather than stationary systems\cite{Theo}.
 It seems  also closely related to the macroscopic temperature
 driving activated proceses in the SGR model\cite{Sollich}.

\vspace{.5cm}

{\bf A Simple Example}

\vspace{.5cm}

For concreteness, let us consider a variation of the standard
mean-field glass model \cite{crso,review}. The variables are $x_i$, $i=1,...,N$, and are 
suject to an equation of motion:
\begin{equation}
m {\ddot x}_i + \frac{\delta E( \bbox{x})}{\delta x_i} + \Omega x_i =
- \epsilon f^{\mbox{`shear'}}_i( \bbox{x}) - f^{\mbox{`friction'}}_i({ \bbox{\dot x}}) 
\label{motion}
\end{equation}

The left hand side is just Newtonian dynamics (with $\Omega$ possibly time-dependent), with a
`glassy' potential which we can take, for example, as:
\begin{equation}
E( \bbox{x}) = \sum J_{ijk} x_i x_j x_k
\end{equation}
where the $J_{ijk}$ is a symmetric tensor of random quenched  variables
of variance $1/N^2$. These terms correspond to the p-spin glass \cite{crso}.
It was realised some ten years ago that this kind of model constitutes 
a mean-field caricature of fragile glass \cite{KTW}, and in particular
 its dynamics 
yields above the glass transition the simplified mode-coupling equations \cite{mocu}.

On the right hand-side of Eqn. (\ref{motion}) we have added two terms that mimic 
granular experiments.
The forces $f_i$  do not derive from a potential,
for example \cite{Cukulepe,Berthier}:
\begin{equation}
f^{\mbox{`shear'}}_i(\bbox{x})= K^i_{jk} x_j x_k
\end{equation}
with $K^i_{jk}$ a {\em non-symmetric} tensor with random elements with variance $1/N$.
They pump energy into the system, and hence play a role similar to shearing. {\em All
our discussion will be restricted to weak driving, i.e.} $\epsilon$ {\em small}.
 For the  friction terms we can take, instead of a linear term $\propto {\dot x}_i$, a more
complicated {\em odd} function  $f^{\mbox{`friction'}}_i=f^{\mbox{`friction'}}(\dot x_i)$.

 These equations for the correlation $C(t,t')=\frac{1}{N}  \sum \langle x_i(t)
 x_i(t')\rangle  $ and response $R(t,t')=\frac{1}{N} \sum \delta \langle x_i(t)
 \rangle / \delta h_i(t')$
 can be exactly solved in the large $N$ limit. One may do so by reducing
 the system to a 
 self-consistent   single-site equation \cite{Sozi}
\begin{eqnarray}
m {\ddot x}+ \Omega x = - \epsilon f^{\mbox{`shear'}}(t)  &-&
f^{\mbox{`friction'}}(\dot x) +  \nonumber \\
&+& 6
\int^t dt' C(t,t') R(t,t') x(t') + \rho(t) 
\label{equa}
\end{eqnarray}
 $f^{\mbox{`shear'}}(t)$ and $\rho(t)$ are independent coloured Gaussian noises that satisfy:
\begin{equation}
<f^{\mbox{`shear'}}(t) f^{\mbox{`shear'}}(t')> = <\rho(t) \rho(t')>=3C^2(t,t') 
\end{equation}
Equation (\ref{equa}) is supplemented by the self-consistency conditions
\begin{equation}
\langle x(t) x(t') \rangle = C(t,t') \;\;\; ; \;\;\; R(t,t')=\frac{\delta  \langle x(t) \rangle}{\delta h(t')}
\label{selfcon} 
\end{equation}
where $h(t)$ is a field that acts additively in (\ref{equa}).

We now perform the usual step of separating `fast' and `slow' functions. Accordingly, we put:
\begin{eqnarray}
C(t,t')=C_F(t,t') + C_S(t,t') \;\;\; &;& \;\;\; R(t,t') = R_F(t,t') + R_S(t,t') \nonumber \\
f^{\mbox{`shear'}}=f^{\mbox{`shear'}}_F+f^{\mbox{`shear'}}_S \;\;\; &;& \;\;\;
\rho = \rho_F + \rho_S
\end{eqnarray}
the additive separation is such that in the region of large time-differences $C_F(t,t')$ and the integral of
$R_F(t,t')$ tend to zero, and the induced noises are now divided into fast and slow:
\begin{eqnarray}
<f^{\mbox{`shear'}}_F(t) f^{\mbox{`shear'}}_F(t')> = <\rho_F(t) \rho_F(t')>&=&3C_F^2(t,t') \nonumber \\
 <f^{\mbox{`shear'}}_S(t) f^{\mbox{`shear'}}_S(t')> = <\rho_S(t) \rho_S(t')>&=&3C_S^2(t,t')
\end{eqnarray}

We can now rewrite the single-site equation (\ref{equa}) in the following way
\begin{eqnarray}
m {\ddot x}+ \Omega x = - \epsilon f^{\mbox{`shear'}}_F(t)  &-&
f^{\mbox{`friction'}}(\dot x)  +  \nonumber \\
&+& 6
\int^t dt' C_F(t,t') R_F(t,t') x(t') + \rho_F(t)  + Z(t) 
\label{fast}
\end{eqnarray}
where:
\begin{equation}
Z(t)= - \epsilon f^{\mbox{`shear'}}_S(t) +  6
\int^t dt' C_S(t,t') R_S(t,t') x(t') + \rho_S(t) 
\label{slow}
\end{equation}
 
Consider equation (\ref{fast}): it describes a single degree of freedom which 
has  nonlinear friction and a (short) memory kernel, plus a {\em slowly varying} field $Z(t)$.
Upon the assumption of large separation of timescales (which will be valid if the system is
weakly sheared and `old'), we can treat $Z(t)$ as adiabatic. However, because of the 
absence of detailed balance, we know that the distribution for fixed $Z$ is {\em non-Gibbsean}.
The fast correlation and response functions will be of the form
$C_F(t,t')= C_F(t-t')$ and  $R_F(t,t')= R_F(t-t')$, but
we cannot say anything in general about their relation. 
The average of $\langle\langle x \rangle \rangle_Z$
 over an interval
of time large compared to the fast relaxation (the range of $C_F$ and $R_F$) is 
a certain function of $Z$:
\begin{equation}
\langle\langle x \rangle \rangle_Z = \frac{\partial F(Z)}{\partial Z}
\end{equation}
which defines the single variable function $F(Z)$.

We now turn to the slow evolution. Because the memory kernels in equation (\ref{slow})
are slowly varying and smooth, we can substitute $x$ by its average 
$\langle\langle x \rangle \rangle_Z$.
We hence have:
\begin{equation}
Z(t)= - \epsilon f^{\mbox{`shear'}}_S(t) +  6
\int^t dt' C_S(t,t') R_S(t,t')  \frac{\partial F(Z)}{\partial Z}(t') + \rho_S(t)  + h^{\mbox{adiab}}(t)
\label{slow1}
\end{equation}
where we have explicitated  the slow component a field acting additively in (\ref{equa}).
The self-consistency equations now read:
\begin{equation}
 C_S(t,t')= 
\langle \frac{\partial F(Z)}{\partial Z}(t)  \frac{\partial F(Z)}{\partial Z}(t') \rangle
\;\;\; ; \;\;\; 
 R_S(t,t')= \frac{\delta }{\delta
h^{\mbox{adiab}}(t')} \langle \frac{\partial F(Z)}{\partial Z}(t) \rangle
\label{selfslow2}
\end{equation}

Equations (\ref{slow1}) and (\ref{selfslow2})  describe the slow part of the evolution.
The only input of the fast equations is through the function $F(Z)$.
 The manner of  solution depends little on the fact that the `fast' evolution is
not thermal, and is  by now  standard. Here I sketch the steps \cite{Le} for
completeness.
Our aim is to show that they admit in the small $\epsilon$ limit  a solution of the form:
\begin{eqnarray}
C_S(t,t') = {\tilde C} \left(\frac{h(t')}{h(t)} \right) \;\;\; ; \;\;\;
R_S(t,t') = \beta^* \frac{\partial}{\partial t'} {\tilde C}\left(\frac{h(t')}{h(t)} \right)
\label{uno}
\end{eqnarray}
where $\beta^*$ is the effective temperature that emerges for the slow dynamics, to be
determined by the matching with the `fast' time-sector.
To show this, we first write $\tau=\ln(h(t))$, $\tau'=\ln(h(t'))$, etc, and put
\begin{eqnarray}
 C_S(t,t') &=& {\hat{C}}(\tau-\tau') \nonumber \\
R_S(t,t') &=&  \beta^*  \frac{\partial}{\partial \tau'}{\hat{C}}(\tau-\tau') \nonumber \\
Z(t) &=&  {\hat{Z}}(\tau)
\label{dos}
\end{eqnarray}
Equations  (\ref{slow1})  and (\ref{selfslow2}) now take the form, in the
$\epsilon \rightarrow 0$ limit:
\begin{equation} 
{\hat{Z}}(\tau)= - \epsilon f^{\mbox{`shear'}}_S(\tau) +  6 \beta^* 
\int^\tau d\tau' {\hat{C}}(\tau-\tau')  {\hat{C}}'(\tau-\tau')
 \frac{\partial F({\hat{Z}})}{\partial {\hat{Z}}}(\tau') + \rho_S(\tau)  
+ h^{\mbox{adiab}}
\label{slow11}
\end{equation}
where we have explicitated  the slow component a field acting additively in (\ref{equa}).
The self-consistency equations  read:
\begin{equation}
 {\hat{C}}(\tau-\tau')= 
\langle \frac{\partial F({\hat{Z}})}{\partial {\hat{Z}}}(\tau) 
 \frac{\partial F({\hat{Z}})}{\partial {\hat{Z}}}(\tau') \rangle
\label{selfslow11}
\end{equation}
and
\begin{equation}
\beta^*  \frac{\partial}{\partial \tau'}{\hat{C}}(\tau-\tau')
 = \frac{\delta  }{\delta
h^{\mbox{adiab}}(\tau')}
\langle \frac{\partial F({\hat{Z}})}{\partial {\hat{Z}}}(\tau) \rangle
\label{selfslow22}
\end{equation}

To prove that (\ref{dos}) is a solution of the system (\ref{slow11}) - (\ref{selfslow22}),
and hence that (\ref{uno}) is a solution of  (\ref{slow1}) and (\ref{selfslow2}) we can proceed
as follows: We introduce an infinite set of auxiliary variables $l_i(\tau)$ and 
a set of dynamical variables $y_i$ satisfying an ordinary Langevin equation with inverse temperature
$\beta^*$:
\begin{equation}
\left[ m_j \frac{d^2}{d\tau^2} + \Gamma_j \frac{d}{d \tau} +
 \Omega_j \right] y_j + l_j(\tau) = \Delta_j y_j + l_j(\tau)= \xi_j(\tau)- 
\frac{\partial F \left( \sum_j A_j y_j \right) }{\partial y_j}
\label{a}
\end{equation}
with
\begin{equation}
\langle \xi_i(\tau) \xi_j(\tau_w) \rangle = 2 T^* \Gamma_j \delta_{ij} \delta(\tau-\tau_w) 
\; .
\label{truc}
\end{equation}
We can now choose the $A_j$ and the $l_j$ such that:
\begin{eqnarray}
\Delta_j^{-1} \sum_j A_j \xi_j(t) &=&  \rho(t) \nonumber \\
\sum_j A^2_j \Delta_j^{-1} &=& {\hat{C}}'(\tau) \Theta (\tau) \nonumber \\
\sum_j A_j \Delta_j^{-1} l_j(\tau) &=& h^{\mbox{adiab}}(\tau)
\end{eqnarray}
and   check that the quantity $\sum_j A_j y_j$ obeys the same equation of motion as $Z(\tau)$.

Because the problem reduced to an  ordinary (not glassy!) Langevin equation, we can  assure
that the system thermalises at temperature $T^*=1/\beta^*$, and hence verify that the
ansatz closes.

We have followed  essentially the same steps as in the treatment of a glassy models coupled
to a `good' heat bath. Here instead of having  two time
scales each with its own temperature we have a temperature associated only with the
low frequency  motion.

\pagebreak

{\bf Conclusions}

\vspace{.5cm}

 It has been recognised for some time that  granular  systems  bear a deep similarity
with  glassy systems  at essentially zero temperature.
In order to introduce some agitation, both tapping and 
shearing have been often introduced.
This certainly makes sand look more like a fluid, but at
 the same time poses the problem
of introducing  (and dissipating) energy in a manner that
 is quite different from that of
a thermal bath.

We have shown  in this paper that, at least within an approximation scheme and for
{\em slowly flowing systems}, one can still introduce
 thermodynamic concepts --- provided
one attempts to apply them only to the low frequency motion.
  
For very small power input, it turns out that the fluctuation-dissipation
temperature of the slow degrees of freedom coincides in
these models with Edwards', defined on the basis of the logarithm of the
number of  stable configurations\cite{jamming}. For stronger driving power, an
analogous definition needs the counting of stable states each composed of
many configurations.

\vspace{.5cm}

{\bf Aknowledgements}

\vspace{.5cm}

I wish to thank Anita Mehta for useful comments.

\vspace{.5cm}

%%%%%%%%%%%%%%%%%%%%%%%%%%%%%%%%%%%%%%%%%%%%%%%%%%%%%%%%%%%%%%%%%%%%%%%
%                    BIBLIOGRAPHY
%%%%%%%%%%%%%%%%%%%%%%%%%%%%%%%%%%%%%%%%%%%%%%%%%%%%%%%%%%%%%%%%%%%%%%%

\end{document}